\documentclass[fleqn,10pt]{wlscirep}
\usepackage[utf8]{inputenc}
\usepackage[T1]{fontenc}

\def\be{\begin{equation}}
\def\ee{\end{equation}}
\def\bea{\begin{eqnarray}}
\def\eea{\end{eqnarray}}

\newcommand{\ket}[1]{|#1\rangle}

\newcommand{\braket}[2]{\langle #1|#2\rangle}

\newcommand{\expectval}[1]{\langle #1\rangle}
\newcommand{\sandwich}[3]{\langle #1|#2|#3\rangle}

\title{Thermal bosons in 3d optical lattices via tensor networks}

\author[1]{Saeed S. Jahromi}
\author[1,2,3,*]{Rom\'an Or\'us}
\affil[1]{Donostia International Physics Center, Paseo Manuel de Lardizabal 4, E-20018 San Sebasti\'an, Spain}
\affil[2]{Ikerbasque Foundation for Science, Maria Diaz de Haro 3, E-48013 Bilbao, Spain}
\affil[3]{Multiverse Computing, Paseo de Miram\'on 170, E-20014 San Sebasti\'an, Spain}

\affil[*]{roman.orus@dipc.org}

\affil[+]{these authors contributed equally to this work}

\begin{abstract}
Ultracold atoms in optical lattices are one of the most promising experimental setups to simulate strongly correlated systems. However, efficient numerical algorithms able to  benchmark experiments at low-temperatures in interesting 3d lattices are lacking. To this aim, here we introduce an efficient tensor network algorithm to accurately simulate thermal states of local Hamiltonians in any infinite lattice, and in any dimension. We apply the method to simulate thermal bosons in optical lattices. In particular, we study the physics of the (soft-core and hard-core) Bose-Hubbard model on the infinite pyrochlore and cubic lattices with unprecedented accuracy. Our technique is therefore an ideal tool to benchmark realistic and interesting optical-lattice experiments.
\end{abstract}
\begin{document}

\flushbottom
\maketitle
%
%

\section*{Introduction}

Recent progress in the theory and experiments of ultracold atoms \cite{Bloch2008,Lewenstein2007} has made them one of the preferred frameworks to implement quantum simulations of strongly correlated systems (SCS) \cite{Bloch2008a,Jaksch2005}, such as quantum magnets \cite{Alet2006,Lacroix2013} and High-Tc superconductors \cite{Kohl2005,Ospelkaus2006,Gunter2006,Stoferle2006}. Moreover, the observation of many fascinating phenomena such as Bose-Einstein condensation (BEC) \cite{Anderson1995,Davis1995,Cornell2002,Ketterle2002}, superfluid-Mott insulator transition \cite{Jaksch2005,Jaksch1999,Greiner2002,Bloch2004,Fisher1989} and BEC-BSC (Bardeen-Cooper-Schrieffer) crossover \cite{Regal2004,Bartenstein2004,Chin2004,Zwierlein2004,Kinast2004,Bourdel2004,Zwierlein2005,Zwierlein2006,Partridge2006} are nowadays possible thanks to on-going experimental efforts. State-of-the-art optical lattices allow to accurately simulate a variety of Hubbard models at low temperatures (nanoKelvin regime) as well as complex bosonic \cite{Jaksch2005} and spin quantum many-body systems \cite{Garcia-Ripoll2004,Santos2004,Duan2003,Dorner2003}.  
 
In parallel to experiments, advances in numerical methods have also allowed a better understanding of quantum matter. Techniques such as quantum Monte Carlo (QMC) \cite{Sandvik2010,Sandvik1991,Sandvik2003,Capponi2014} and tensor network (TN) methods \cite{Orus2014,Orus2014a,Orus2019,Ran2017,Biamonte2017,Verstraete2008} have played a major role in this respect, not only by allowing to benchmark experiments, but also  by motivating new lines of research. In particular, QMC has been so far the main benchmark algorithm for validating ultracold atom experiments at finite-temperature \cite{Mahmud2011,Capogrosso-Sansone2007,Capogrosso-Sansone2008,Fang2011,Zhou2009,Freericks1996,Rigol2009}. Besides, TN algorithms such as Density Matrix Renormalization Group (DMRG) \cite{White1993,White2004} and those based on Projected Entangled-Pair States (PEPS) \cite{Verstraete2006,Verstraete2008,Orus2014,Orus2014a} have been remarkably successful in studying SCS both at zero- \cite{Corboz2014a,Corboz2012a,Corboz2013,Corboz2014,Jahromi2018,Jahromi2018a,Schmoll2019,Sadrzadeh2016,Jahromi2020,Jahromi2019} and finite-temperature \cite{Wietek2019,Kshetrimayum2019,Qu2019,Ran2019,Czarnik2012,Czarnik2015a,Czarnik2015,Zwolak2004,Kshetrimayum2017,Verstraete2004a}, in 1d and 2d. In spite of its success, QMC is plagued with the sign-problem, which limits its application for fermionic and frustrated systems. Likewise, state-of-the-art TN methods are highly lattice-dependent, and mostly tailored to target low-temperature properties \cite{Corboz2012a,Corboz2013,Jahromi2018,Jahromi2018a,Schmoll2019}, with advances for thermal states limited mostly to 1d and, only recently, also to 2d \cite{Kshetrimayum2019,Qu2019,Czarnik2015a}. It is therefore essential to develop new numerical techniques able to overcome all these limitations, and which are sufficiently accurate in order to benchmark the complex experiments in state-of-the-art optical lattices. 

Considering the above, here we introduce an accurate and highly-efficient TN method for the simulation in the thermodynamic limit of thermal properties of local Hamiltonians in {\it any} dimension and lattice geometry. We call this the \emph{thermal graph-based PEPS (TgPEPS) algorithm}. To show how powerful our method is, we use it to efficiently  simulate the physics of the 3d Bose-Hubbard (BH) model in the pyrochlore (see Fig.~\ref{Fig:lattices}(a)) and cubic lattices, including the zero- and finite-temperature superfluid-Mott insulator transitions as well as the full phase diagrams and  critical properties, with unprecedented accuracy.

\section*{The TgPEPS Method}
\begin{figure}
\centerline{\includegraphics[width=10cm]{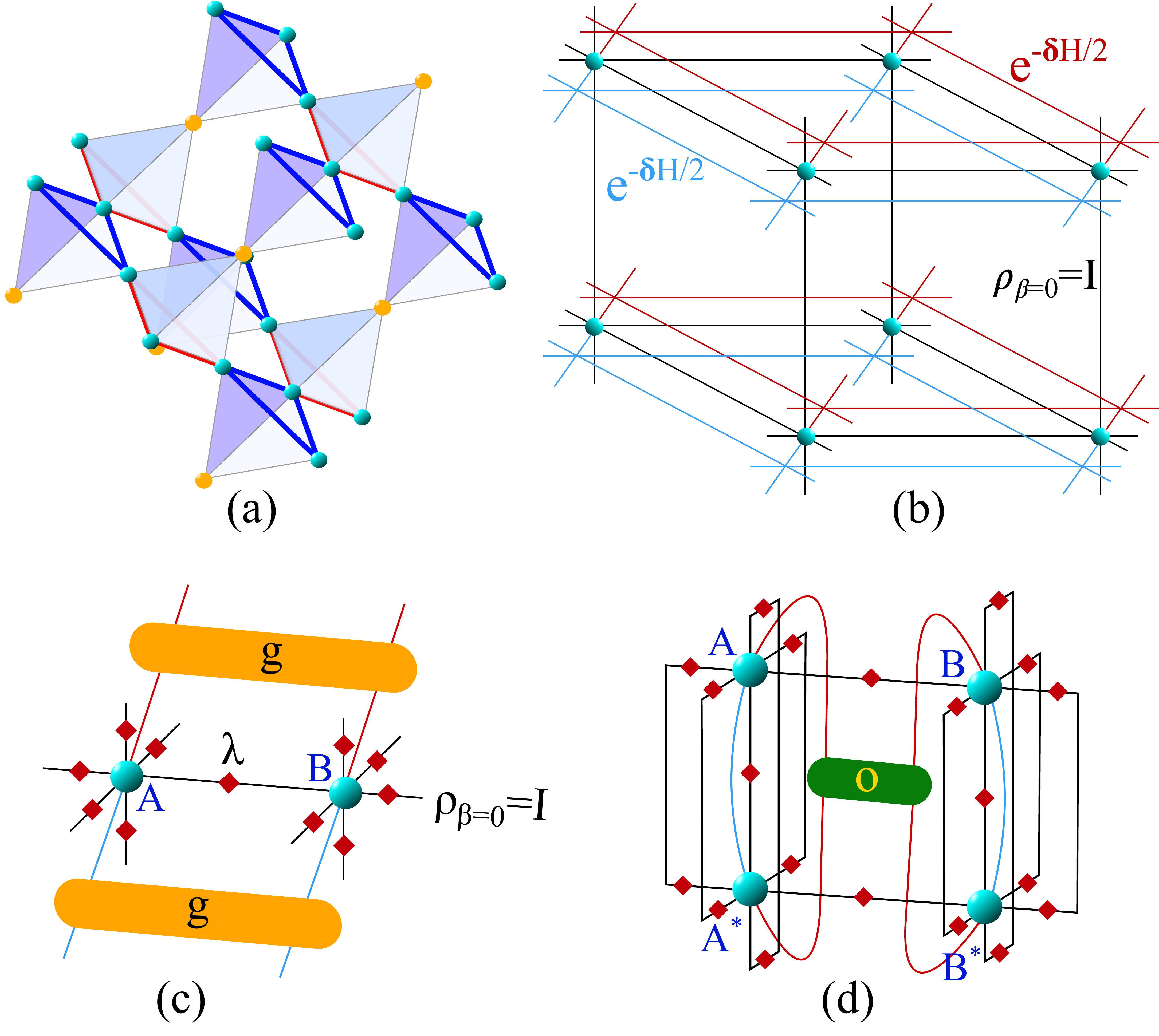}}
\caption{(Color online) (a) Pyrochlore lattice of corner-sharing tetrahedrons. (b) PEPO for the cubic lattice. In this tensor network diagram, shapes are tensors, lines are indices, and connected lines are contracted common indices. Red and blue indices at every tensor correspond to the local bra and ket degrees of freedom, respectively. (c) Action of the Suzuki-Trotter gate $g$ on both ket and bra indices of two nearest-neighbouring sites, for the PEPO tensors of the cubic lattice. Small red tensors $\lambda$ correspond to the matrices of singular values obtained from the simple update \cite{Jiang2008}. (d) Thermal expectation value of a local two-body operator $O$ in the cubic lattice, with a mean-field approximation of the environment (see main text for more details).}
\label{Fig:lattices}
\end{figure}

Traditionally, TN methods for zero-temperature properties target ground states of local Hamiltonians by, e.g., variational optimization \cite{Verstraete2008,Corboz2016} and imaginary-time evolution (ITE) \cite{Vidal2004,Orus2008,Vidal2007}. However, at finite-temperature we target the  thermal density matrix (TDM) of Hamiltonian $H$, i.e., $\rho=e^{-\beta H}$, $\beta=1/T$ being the inverse temperature. To approximate this state, one typically evolves in imaginary-time for a time $\beta/2$ both the bra and ket degrees of freedom starting from the infinite-temperature state, i.e., $\rho = e^{-\beta H/2} \cdot {\mathbb I} \cdot e^{-\beta H/2}$ \cite{Zwolak2004, Verstraete2008, Kshetrimayum2019,Qu2019,Czarnik2012}. In TN language, the TDM of a system on a lattice with coordination number $z$ can be described by a Projected Entangled-Pair Operator (PEPO) \cite{Orus2014}. The typical tensor describing this PEPO is of the type  $T_{i,j,\alpha_1,\cdots,\alpha_z}$, with $i,j =1, \ldots, p$ and $\alpha_k = 1, \ldots, D$, $p$ being the dimension of the local Hilbert space and $D$ the bond dimension of the bond indices, which in turn controls the amount of correlations (classical \emph{and} quantum) that can be handled by the ansatz, see Fig.~\ref{Fig:lattices}(b) for the example of the cubic lattice. 

Let us consider, without loss of generality, the case of a Hamiltonian with nearest-neighbour interactions, $H=\sum_{\langle i,j \rangle} h_{ij}$.  The TDM can be sliced into $m$ ITE steps using the Suzuki-Trotter decomposition, i.e,
\be
\rho\approx \left(\prod_{\langle i,j \rangle} e^{-\delta h_{ij}}\right)^{m/2} \cdot {\mathbb I} \cdot  \left(\prod_{\langle i,j \rangle} e^{-\delta h_{ij}}\right)^{m/2},
\ee
where $\delta \ll 1$ and $m  \cdot \delta =\beta$. The thermal density matrix of the system at inverse temperature $\beta$ is then obtained after $m$ successive applications of the gates $g \equiv e^{-\delta h_{ij}/2}$ on the corresponding links of the lattice, as shown in Fig.~\ref{Fig:lattices}(b,c) for the cubic lattice. 

After applying a gate on the PEPO, the bond dimension of the index connecting the local sites grows from $D$ to $p^2D$. To truncate it back to its original size one can use a variety of methods, including the so-called ``simple" update (SU) \cite{Jiang2008,Corboz2010a,Jahromi2019}, or the more involved ``fast-full" \cite{Phien2015} and ``full" updates (FFU, FU) \cite{Corboz2014,Corboz2014a,Jahromi2018a}. These algorithms differ in the way the handle the correlations around the link to be truncated. While the SU handles these correlations by a mean-field-like approach and truncates the index directly with a Singular Value Decomposition (SVD), FFU and FU approximate the full effect of the environment via powerful techniques such as Tensor Renormalization Group (TRG) \cite{Levin2007, Gu2008} and Corner Transfer Matrix Renormalization Group (CTMRG) \cite{Nishino1996,Orus2009,Orus2012,Corboz2014a,Corboz2010a}. The price to pay, though, is that FFU and FU are computationally much more expensive than SU. Besides, application of the FU and FFU algorithms to lattices with higher dimensionality and connectivity is highly non-trivial. However, for thermal systems with a lot of connectivity (such as higher-dimensional systems), the mean-field-like approximation of the environment in the SU is actually good, in turn making the SU a quite accurate option in these situations. And this are good news, because the computational cost of the FFU and FU algorithms is really high for thermal high-dimensional systems.  

The TgPEPS algorithm thus targets both the geometrical challenges \emph{and} the efficient approximation of the environment, in turn providing a universal algorithm for thermal states applicable to any local Hamiltonian in any dimension. Geometrical aspects of the network are encoded in the so-called {\it structure-matrix} (SM), first proposed by the authors in Ref.~\cite{Jahromi2019}. Each column of the SM corresponds to one of the links of the lattice and contains all the details about the neighbouring tensors, their interconnecting indices, and their bond dimensions. Thanks to this one can fully automatize the TN update by looping over the columns of the SM in a very systematic way, without the burden of complications due to geometry (see Refs.\cite{Jahromi2019} for detailed discussions and examples of SM for different lattices, as well as the Appendix for more information). 

\begin{figure}
\centerline{\includegraphics[width=16cm]{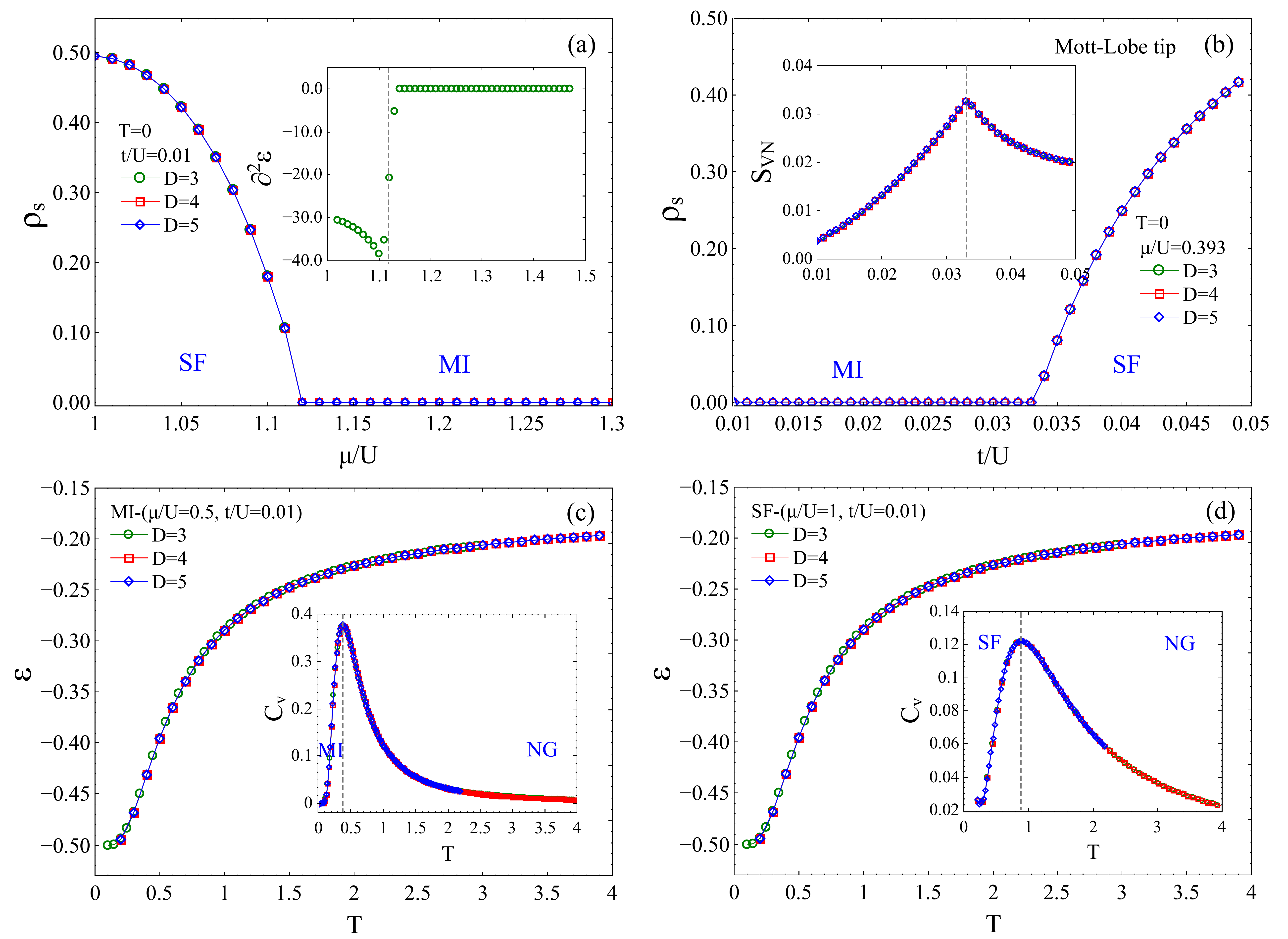}}
\caption{(Color online) TN results for the soft-core ($n=2$) BH model on the $3d$ cubic lattice. (a) Superfluid density, $\rho_s$, for $t/U=0.01$ and $T=0$. $\rho_s$ is finite in the superfluid (SF) phase and vanishes in the Mott-insulating (MI) phase. The inset shows the second derivative of the ground-state energy per-site, $\varepsilon_0$, revealing the 2nd-order nature of the SF-MI quantum phase transition at $T=0$. (b) The $\rho_s$ at the tip of the first Mott lobe at $\mu/U=0.393$, and also at $T=0$. The inset shows the Von-Neumann entropy for bipartition of the lattice on a link (obtained from singular values of $\lambda$ bond environment matrices) signalling the transition at $(t/U)^c=0.0331$. (c) Energy per site $\varepsilon$ as a function of $T$ in the MI phase. Specific heat $C_v$ is shown in the inset. (d) Energy per site $\varepsilon$ as a function of $T$ in the SF phase. Specific heat $C_v$ is shown in the inset.}
\label{Fig:EnCV}
\end{figure}

Moreover, TgPEPS makes explicit use of the SU which, as explained above, is well-suited for finite-temperature systems of high dimensionality. Extensions using the FFU and FU (on simple lattice structures) are of course possible, but we found that the SU already provides remarkable accuracy in the considered regimes. Our thermal SU algorithm is based on an iterative ITE algorithm that updates all links of the lattice in each iteration by incorporating the mean-field-like environment (see Appendix for details). Thanks to the  SM, the algorithm can also be efficiently adapted to local Hamiltonians on any lattice geometry and dimension \cite{Jahromi2019}. The algorithm can also be enhanced by using a local gauge-fixing of the tensors \cite{Jahromi2019,Evenbly2018,Ran2012a}.

After obtaining the PEPO approximation of the thermal state $\rho$, expectation values of local operators $\langle O \rangle_\beta = {\rm Tr}(\rho(\beta) O)/Z(\beta)$ where $Z(\beta)$ is the partition function of the system. The density operator $\rho(\beta)$ is the product of two half density operators $\rho(\beta/2)$, i.e,
\be
\rho(\beta)=\rho(\beta/2)\rho(\beta/2)=\rho^\dagger(\beta/2)\rho(\beta/2).
\ee   
Using the "Choi's" isomorphism \cite{Zwolak2004,Kshetrimayum2017}, the density operator $\rho(\beta/2)$ can be replaced by its vectorized form $\ket{\rho(\beta/2)}$ in the product space of the ket and bra. In this vectorized representation, the partition function then becomes an inner product of $\ket{\rho(\beta/2)}$ and its vector conjugate, namely
\be
Z(\beta)={\rm Tr}\rho(\beta)=\braket{\rho(\beta/2)}{\rho(\beta/2)},
\ee
and the un-normalized expectation value of local operators reads
\be
\expectval{O}_\beta=\sandwich{\rho(\beta/2)}{O}{\rho(\beta/2)}.
\label{eq:expectval}
\ee
The tensor network representation of Eq.~\eqref{eq:expectval} for a two-body operator is demonstrated in Fig.~\ref{Fig:lattices}(d) using the mean-field approximation to the environment which, as we argued before, works well in the considered regimes \footnote{Notice also that, unlike at zero temperature, here we do not target variational approximations of ground-state energies.}. In the end, our approach allows us to push the simulation of thermal 3d models in any geometry to very large dimensions with a very cheap computational cost of $O(p^2D^z)$. 

\section*{Numerical results}

 \begin{figure}
\centerline{\includegraphics[width=14cm]{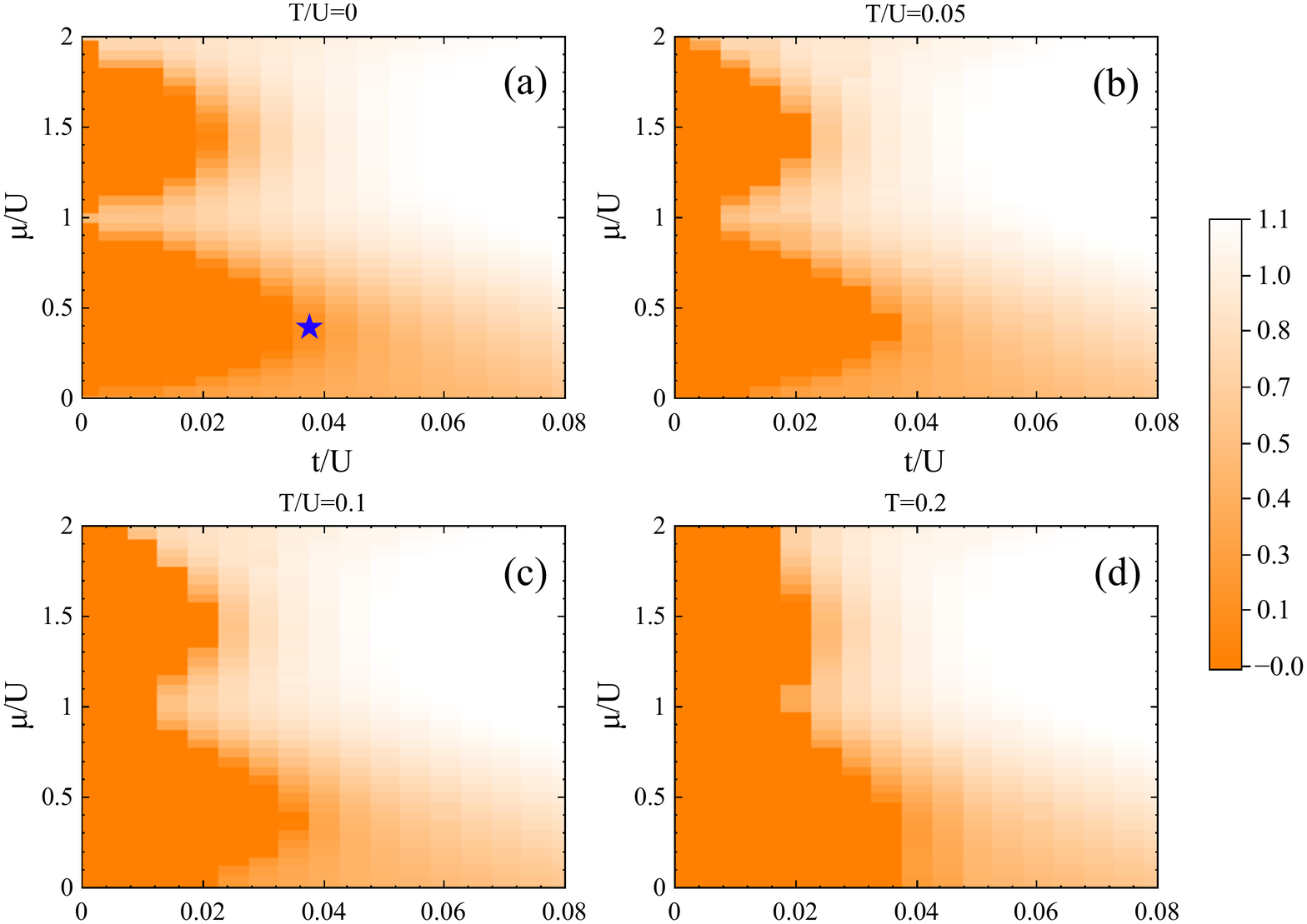}}
\caption{(Color online) Contour-plot of the superfluid density $\rho_s$ for the soft-core ($n_{oc}=2$) BH model on the cubic lattice at (a) $T=0$, (b) $T=0.05$, (c) $T=0.1$ and (d) $T=0.2$. The blue star in (a) locates the tip of the first Mott lobe. The data points vary from zero in the dark region (MI) to one in the light region (SF). Increasing  $T$ shrinks the SF regions.}
\label{Fig:phasediag}
\end{figure}

To prove the validity of our approach, we apply the TgPEPS technique to the Bose-Hubbard (BH) model \cite{Fisher1989} in 3d, for which we study the low-temperature phase diagram in the cubic and pyrochlore lattices, with maximum occupation number $n_{oc} = 2$ (soft-core) and $n_{oc} = 1$ (hard-core) respectively. The pyrochlore lattice is one of the most challenging structures from the point of view of TN simulations. Besides, the hard- and soft-core cases have different physical dimensions, i.e., $p=2$ and $p=3$, respectively resulting in different computational complexity. Our choice of these lattices and occupation numbers was to seriously challenge our algorithm and assess the stability and accuracy of the simulations for different physical systems and lattice geometries.    

The generic Hamiltonian of the Bose-Hubbard model is given by 
\bea 
H=-t\sum_{\langle ij\rangle} (a^{\dagger}_i a_j+a^{\dagger}_j a_i)+\frac{U}{2}\sum_i n_i(n_i-1)-\mu\sum_i n_i,\nonumber \\  
\eea 
where $a^{\dagger}$ ($a$) are the bosonic creation (annihilation) operators, $n = a^{\dagger} a$ is the particle number operator, $t$ is the hopping rate between nearest-neighbour sites, $U$ is the on-site repulsive interaction, and $\mu$ the chemical potential. At zero-temperature $T=0$, in the extreme regime where $t/U\ll1 $ the repulsive interaction is very strong and only one particle per site is allowed, resulting in a Mott insulating (MI) phase. Complementarily, in the $t/U\gg1$ regime the bosons are highly delocalized, and the system is in a  coherent superfluid (SF) phase. One therefore expects a quantum phase transition (QPT) between these two regimes.        

Let us first consider the $T=0$ properties for the soft-core ($n_{oc}=2$) BH model on the cubic lattice, which we can compute accurately using our technique from Ref.~\cite{Jahromi2019}. Fig.~\ref{Fig:EnCV}(a) shows the $T=0$ SF-MI transition at $t/U=0.01$. This is captured by the superfluid order parameter $\rho_s=|\expectval{a_i}|^2$ which is nonzero in the SF phase and zero in the MI phase. The QPT takes place at $(\mu/U)^c\approx 1.119$. The discontinuity in the second derivative of the ground state energy per site (inset of Fig.~\ref{Fig:EnCV}(a)) confirms that this transition is second-order. We have further mapped out the full $T=0$ phase diagram in the $t/U$-$\mu/U$ plane up to $n_{oc}=2$ in Fig.~\ref{Fig:phasediag}(a). The figure depicts the contour-plot of the superfluid density, which is zero in the two Mott lobes (dark regions). 

\begin{figure}
\centerline{\includegraphics[width=12cm]{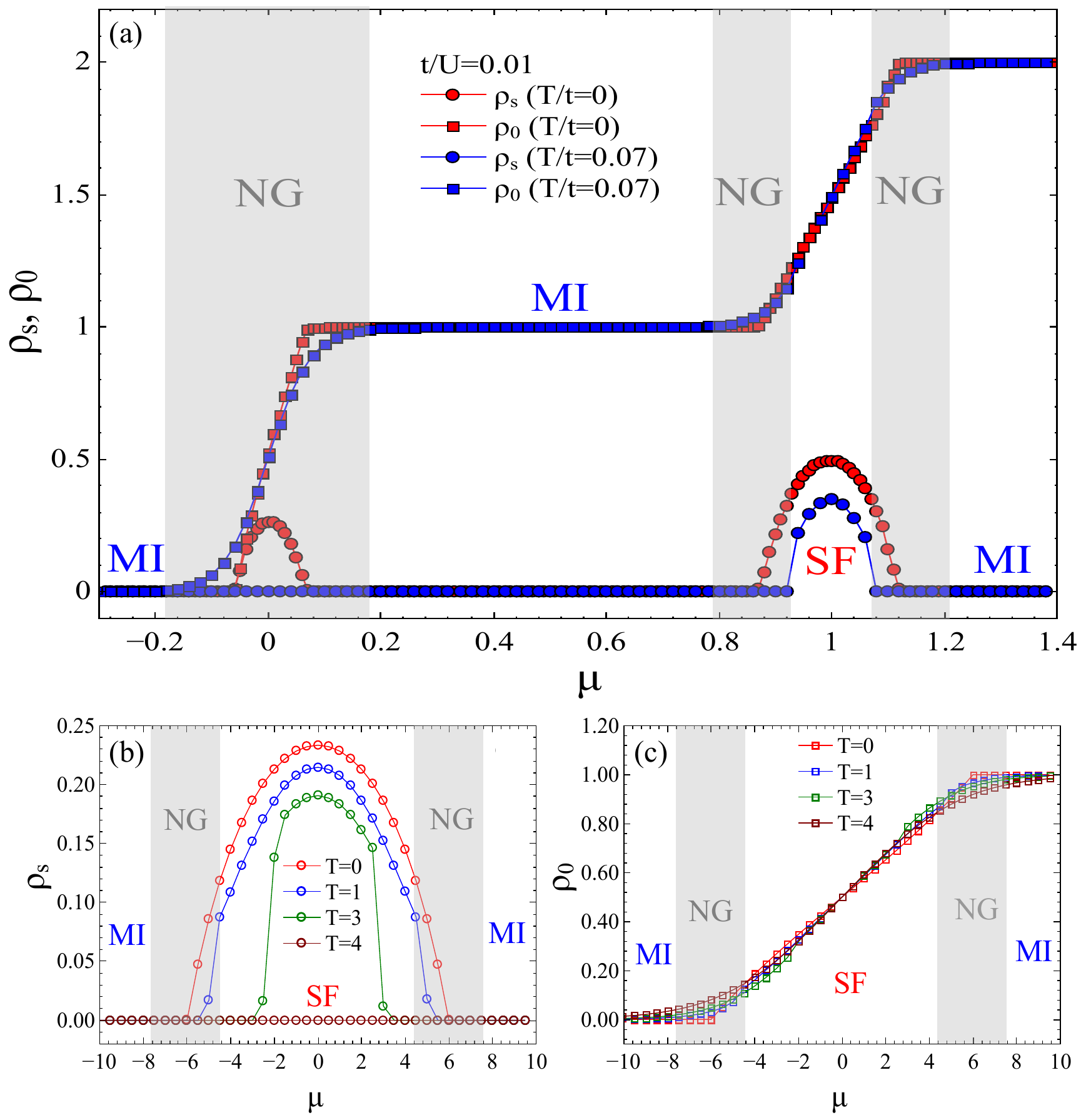}}
\caption{(Color online) (a) Superfluid density $\rho_s$, and particle density $\rho_0$, for the 3d BH model on the cubic lattice for $t/U=0.01$, $T/t=0$ and $T/t=0.07$. While the SF phase is characterized by non-zero $\rho_s$, both MI and NG (which exists for $T>0$) phases have zero $\rho_s$. However, the MI phase has integer $\rho_0$ whereas NG phases (shaded grey regions) have non-integer $\rho_0$. (b) $\rho_s$ and (c) $\rho_0$ for the hard-core ($n_{oc} = 1$) BH model on the pyrochlore lattice for $t=1, U=0$. Both plots show how $\rho_s$ and $\rho_0$ evolve when increasing the temperature, eventually having a conventional gas phase at large $T$. The shaded grey regions are the NG phases emerging at $T=1$ (we use $T$ instead of $T/t$ since $t =1$ in this case).}
\label{Fig:hardcore}
\end{figure}

It is already known that there exist two different types of transitions between the Mott insulator and superfluid regions of the Bose-Hubbard phase diagram. The trivial phase transition occurs when the MI-SF phase boundary is crossed at fixed $t/U$ and a critical phase transition at fixed integer density when the phase boundary is crossed at fixed $\mu/U$ such as the tip of the Mott lobes. While the former transition can be described by the physics of a weakly interacting Bose gas and trivial (mean-field) universality class, the latter transition which occurs at relatively large $t/U$ is described by highly fluctuating delocalized bosons with emerging particle-hole symmetry and relativistic dispersion \cite{Sanders2019,Capogrosso-Sansone2007,Capogrosso-Sansone2008} that belongs to the (d+1)-XY universality class \cite{Fisher1989,Capogrosso-Sansone2007}. Our accurate analysis shows that the tip of the first lobe (blue star in Fig.~\ref{Fig:phasediag}(a)) is located at $\mu/U\approx0.393$, in excellent agreement with previous studies \cite{Sanders2019,Capogrosso-Sansone2007,Capogrosso-Sansone2008}. We find that the QPT occurs at $(t/U)^c\approx 0.0331$, which is well detected by $\rho_s$ as well as by the one-site Von-Neumann entanglement entropy shown in Fig.~\ref{Fig:EnCV}(b). On the 3d cubic lattice, the QPT at the tip of the lobe is in the four-dimensional XY universality class \cite{Sanders2019,Capogrosso-Sansone2007} which for $d>3$ is equivalent to mean-field universality class hence, is trivial. 
 
Next, we study the thermal properties of the BH model. In experiments with ultracold gases in optical lattices, bosonic atoms are cooled down to the nanoKelvin regime by efficient techniques such as laser cooling \cite{Metcalf2004,Phillips1998} and the desired quantum states are engineered by inducing quantum correlations  between the atoms by microwave pulses. Increasing the temperature of the system destroys quantum correlations due to the extra kinetic energy imposed on the atoms by thermal fluctuations, and eventually the system will end up in a normal gas (NG) phase at large $T$ \cite{Mahmud2011}. A thermal phase transition (TPT) is therefore expected between the underlying quantum state and the NG phase, depending both on $T$ as well as the couplings of the Hamiltonian. 

We further use TgPEPS to provide a good insight into the stability of the Mott and superfluid phases at finite-$T$. For this, we first dive deeply into the MI and SF regions of the zero-$T$ phase diagram computed previously, and then ramp up the temperature from zero up to large $T$. To this end, we fixed  $t/U=0.01$ and computed the TDM of the BH model in the cubic lattice for $\mu/U=0.5$ and $\mu/U=1.0$, which respectively correspond to the MI and SF phases. Fig.~\ref{Fig:EnCV}(c,d) show the energy $\varepsilon$ and specific heat $C=\partial\varepsilon / \partial T$ of the BH model on the cubic lattice versus $T$ in the MI and SF phases. The ideal MI phase only exist at $T=0$. However, at very low-temperature some Mott-like features still persist in the system but with a small finite compressibility. The boundary between the MI and NG phases is therefore a thermal crossover not a thermal phase transition. Our analysis shows that the MI-NG crossover occurs at $T'\approx0.38$ as can be captured by the peak of the specific heat in Fig.~\ref{Fig:EnCV}(c). On the other hand, the superfluid phase can exist even at finite-temperature and. The SF-NG phase boundary is therefore a true thermal phase transition which for the example couplings of our choice is located at $T_c\approx0.87$ (see the peak of $C_v$ in Fig.~\ref{Fig:EnCV}(d)). Our TN thermometry analysis thus, indicates that both MI and SF states are stable only at very low temperatures which evaporate to a normal gas for $T/t\gtrapprox1$.

Additionally, we have mapped the finite-$T$ phase diagram of the model in the $t/U$-$\mu/U$ plane. Figs.~\ref{Fig:phasediag}(b-d) illustrate the superfluid density $\rho_s$ for $T=0.05, 0.1, 0.2$, revealing how the thermal fluctuations shrink the SF region. The distinction between the MI and NG phases (dark regions) is not visible in the $\rho_s$ plot because $\rho_s=0$ for both. However, one can distinguish them  by observing the particle density $\rho_0=\expectval{a^{\dagger}_i a_i}$, which is shown in Fig.~\ref{Fig:hardcore}(a): it holds integer values in the MI phase, and non-integer values in the NG phase. 

Finally, we used TgPEPS to study the hard-core BH model on the pyrochlore lattice at finite temperature. This lattice is well-known for being responsible for interesting effects in frustrated quantum magnets and Kitaev materials. Fig.~\ref{Fig:hardcore}(b,c) shows the superfluid density and the particle density of this model for various temperatures. The phase boundaries between SF and NG and between MI and NG can clearly be identified from the curves in both plots.

Let us note that in principle there is no limit on the choice of occupation number, $n_{oc}$, in our TN simulation of the BH model. However, one should note that by increasing the $n_{oc}$, the computational cost of the simulation is increased as well. Lets us further stress that there is no soft-core cut-off $n_{oc}$ in our algorithms for simulating the physics of each lobe and simply by fixing the proper physical dimension to $p=n_{oc}+1$ (including the unoccupied sites) for each lobe, one obtains the correct physics for all lobs with occupation $n\leq n_{oc}$.

\section*{Conclusions and outlook}

In this paper we introduced TgPEPS, an efficient TN algorithm for finite-temperature simulation of strongly correlated systems in the thermodynamic limit for any lattice and in any spatial dimension. The method follows the ideas that we introduced in Ref.~\cite{Jahromi2019}, extending them to the finite-temperature case, and allowing for very efficient and accurate simulations of thermal systems with large connectivity. We benchmarked the method by computing the zero- and finite-temperature of the 3d Bose-Hubbard model in the pyrochlore and cubic lattices, with unprecedented accuracy. We believe that the TgPEPS algorithm can serve an essential tool for both benchmarking experiments with ultracold atoms in optical lattices with complex geometric structures. While the TgPEPS algorithm presented in this study has been developed for infinite systems, the optical lattice experiments are performed on finite lattice. On should therefore, be careful when benchmarking finite-lattice experiments. Nevertheless, the TgPEPS algorithm can be adopted to finite systems with minor modifications. 

Let us further point out that while the method has been proven to be very successful for simulating local lattice Hamiltonians, some of the longer-range correlations which go beyond the nearest-neighbours may not be fully captured due to the simple mean-field-like environments that have been used for calculating the expectation values and correlators. Besides, after a detailed investigation of the TgPEPS algorithms for simulating some highly entangled systems such as the Kitaev spin liquids, we found out that the algorithms have difficulties in convergence at very low-temperature regime below $T<10^{-3}$ due to the proximity of thermal states to the $T=0$ ground state \cite{Czarnik2019,Jahromi2020a}.  
 
Last but not least, for the lattice geometries that can be adopted to renormalization techniques, such as TRG and CTMRG, the thermal states obtained from TgPEPS method can be used to approximate the full environment from which one can calculate variational energies and expectation values. One can therefore increase the accuracy of the simulations by capturing longer-range correlations in the system. Finally, let us point out that the TgPEPS algorithm can also be applied to simulate fermionic atoms on optical lattices \cite{Corboz2010}, which we will address in the future extensions of this work. All in all, we think that our method will become a very helpful tool in the discovery of new exotic phases of quantum matter.

\section*{Appendix}

\section*{Structure-Matrix for the cubic and pyrochlore lattices}

\begin{figure}[h]
	\centerline{\includegraphics[width=4cm]{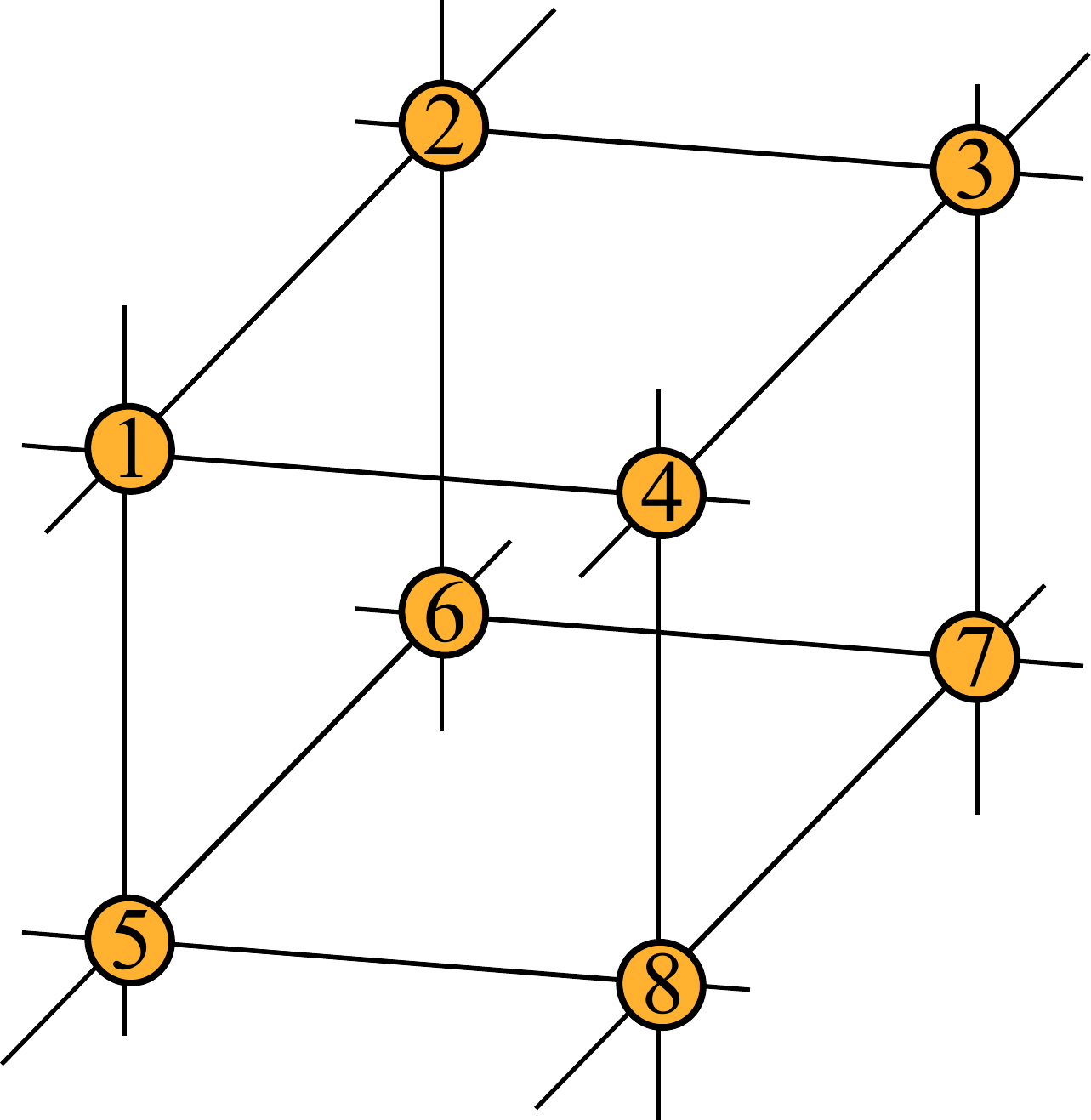}}
	\caption{(Color online) The infinite $3d$ cubic lattice with an $8$-site unit-cell. The numbers at vertices label the graph nodes in the unit-cell.}
	\label{Fig:cubic}
\end{figure}

\begin{figure}[h]
	\centerline{\includegraphics[width=10cm]{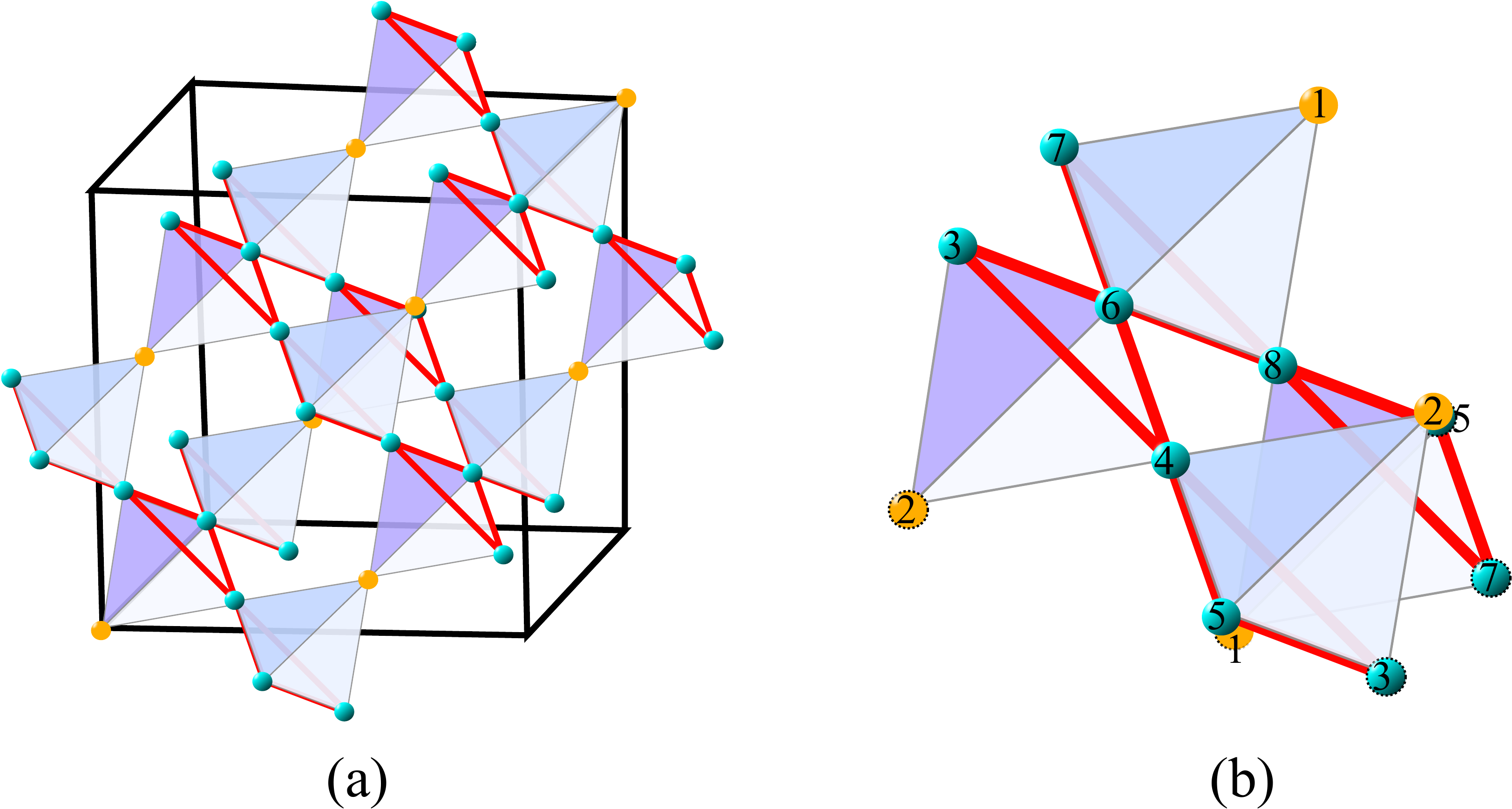}}
	\caption{(Color online) (a) The infinite $3d$ pyrochlore lattice composed of up and down tetrahedrons. (b) The $8$-site unit-cell of the pyrochlore lattice. The numbers represent the labelings of vertices (graph nodes) in the unit-cell.}
	\label{Fig:pyrochlor}
\end{figure}

The structure-matrix (SM) codifies the connectivity information of the tensor network (TN) corresponding to a given lattice geometry. More specifically, the connectivity information of two neighbouring tensors along their shared edges are stored in the columns of the SM. The explicit SMs that we used in our calculations for TN simulation of the cubic (see Fig.~\ref{Fig:cubic}) and pyrochlore lattices (see Fig.~\ref{Fig:pyrochlor}) are provided in Table \ref{tab:smcubic} and \ref{tab:smpyrochlore}, respectively for translationally invariant unit-cells of $8$-sites. For example, according to the second column of Table \ref{tab:smcubic}, the edge $E_3$ connects the bond matrix $\lambda_3$ and the dimensions $4$ and $2$ of tensors $T_1$ and $T_4$, respectively. Thanks to this information, the algorithm can automatically identify the links and tensors where two-body gates are applied, and implement a simple update. This is done by looping over the columns of the SM and systematically updating the iPEPS tensors along their corresponding edges, which can now be done automatically and regardless of the underlying lattice. 

Last but not least, the non-zero elements of the SM at each row start from $2$ which is due to the fact that the first two dimensions ($0,1$) of the tensors $T_i$ in our notation corresponds to the physical bonds of ket and bra, respectively, and play no role in the connectivity of the underlying TN.

\begin{table}[h]
\caption{SM of a $3d$ cubic lattice with an $8$-site unit-cell.}
\centering
\resizebox{\textwidth}{!}{\begin{tabular}{l|llllllllllllllllllllllllll}
  \hline
 & $E_1$ & $E_2$ & $E_3$ & $E_4$ & $E_5$  & $E_6$ & $E_7$ & $E_8$ & $E_9$ & $E_{10}$ & $E_{11}$ & $E_{12}$ & $E_{13}$ & $E_{14}$  & $E_{15}$ & $E_{16}$ & $E_{17}$ & $E_{18}$ & $E_{19}$ & $E_{20}$ & $E_{21}$ & $E_{22}$ & $E_{23}$  & $E_{24}$ \\ 
  \hline
$T_1$     & 2 & 3 & 4 & 5 & 6 & 7 & 0 & 0 & 0 & 0 & 0 & 0 & 0 & 0 & 0 & 0 & 0 & 0 & 0 & 0 & 0 & 0 & 0 & 0  \\
		$T_2$     & 2 & 3 & 0 & 0 & 0 & 0 & 4 & 5 & 6 & 7 & 0 & 0 & 0 & 0 & 0 & 0 & 0 & 0 & 0 & 0 & 0 & 0 & 0 & 0  \\
		$T_3$     & 0 & 0 & 0 & 0 & 0 & 0 & 2 & 3 & 0 & 0 & 4 & 5 & 6 & 7 & 0 & 0 & 0 & 0 & 0 & 0 & 0 & 0 & 0 & 0  \\
		$T_4$     & 0 & 0 & 2 & 3 & 0 & 0 & 0 & 0 & 0 & 0 & 4 & 5 & 0 & 0 & 6 & 7 & 0 & 0 & 0 & 0 & 0 & 0 & 0 & 0  \\
		$T_5$     & 0 & 0 & 0 & 0 & 2 & 3 & 0 & 0 & 0 & 0 & 0 & 0 & 0 & 0 & 0 & 0 & 4 & 5 & 6 & 7 & 0 & 0 & 0 & 0  \\
		$T_6$     & 0 & 0 & 0 & 0 & 0 & 0 & 0 & 0 & 2 & 3 & 0 & 0 & 0 & 0 & 0 & 0 & 4 & 5 & 0 & 0 & 6 & 7 & 0 & 0  \\
		$T_7$     & 0 & 0 & 0 & 0 & 0 & 0 & 0 & 0 & 0 & 0 & 0 & 0 & 2 & 3 & 0 & 0 & 0 & 0 & 0 & 0 & 4 & 5 & 6 & 7  \\
		$T_8$     & 0 & 0 & 0 & 0 & 0 & 0 & 0 & 0 & 0 & 0 & 0 & 0 & 0 & 0 & 2 & 3 & 0 & 0 & 4 & 5 & 0 & 0 & 6 & 7 \\
		  \hline 
\end{tabular}}
\label{tab:smcubic} 
\end{table}

\begin{table}[h]
\caption{ SM of a $3d$ pyrochlore lattice with an $8$-site unit-cell}
\centering
\resizebox{\textwidth}{!}{\begin{tabular}{l|llllllllllllllllllllllllll}
  \hline
 & $E_1$ & $E_2$ & $E_3$ & $E_4$ & $E_5$  & $E_6$ & $E_7$ & $E_8$ & $E_9$ & $E_{10}$ & $E_{11}$ & $E_{12}$ & $E_{13}$ & $E_{14}$  & $E_{15}$ & $E_{16}$ & $E_{17}$ & $E_{18}$ & $E_{19}$ & $E_{20}$ & $E_{21}$ & $E_{22}$ & $E_{23}$  & $E_{24}$ \\ 
  \hline
$T_1$     & 2 & 3 & 4 & 5 & 6 & 7 & 0 & 0 & 0 & 0 & 0 & 0 & 0 & 0 & 0 & 0 & 0 & 0 & 0 & 0 & 0 & 0 & 0 & 0  \\
		$T_2$     & 0 & 0 & 0 & 0 & 0 & 0 & 2 & 3 & 4 & 5 & 6 & 7 & 0 & 0 & 0 & 0 & 0 & 0 & 0 & 0 & 0 & 0 & 0 & 0  \\
		$T_3$     & 0 & 0 & 0 & 0 & 0 & 0 & 2 & 3 & 0 & 0 & 0 & 0 & 4 & 5 & 6 & 7 & 0 & 0 & 0 & 0 & 0 & 0 & 0 & 0  \\
		$T_4$     & 0 & 0 & 0 & 0 & 0 & 0 & 0 & 0 & 2 & 3 & 0 & 0 & 4 & 5 & 0 & 0 & 6 & 7 & 0 & 0 & 0 & 0 & 0 & 0  \\
		$T_5$     & 2 & 0 & 0 & 0 & 0 & 0 & 0 & 0 & 0 & 0 & 3 & 0 & 0 & 0 & 4 & 0 & 5 & 0 & 6 & 7 & 0 & 0 & 0 & 0  \\
		$T_6$     & 0 & 2 & 0 & 0 & 0 & 0 & 0 & 0 & 0 & 0 & 0 & 3 & 0 & 0 & 0 & 4 & 0 & 5 & 0 & 0 & 6 & 7 & 0 & 0  \\
		$T_7$     & 0 & 0 & 2 & 3 & 0 & 0 & 0 & 0 & 0 & 0 & 0 & 0 & 0 & 0 & 0 & 0 & 0 & 0 & 4 & 0 & 5 & 0 & 6 & 7  \\
		$T_8$     & 0 & 0 & 0 & 0 & 2 & 3 & 0 & 0 & 0 & 0 & 0 & 0 & 0 & 0 & 0 & 0 & 0 & 0 & 0 & 4 & 0 & 5 & 6 & 7  \\
		  \hline 
\end{tabular}}
\label{tab:smpyrochlore} 
\end{table}

\section*{The TgPEPS algorithm for any infinite lattice}

\begin{figure*}
	\centerline{\includegraphics[width=16cm]{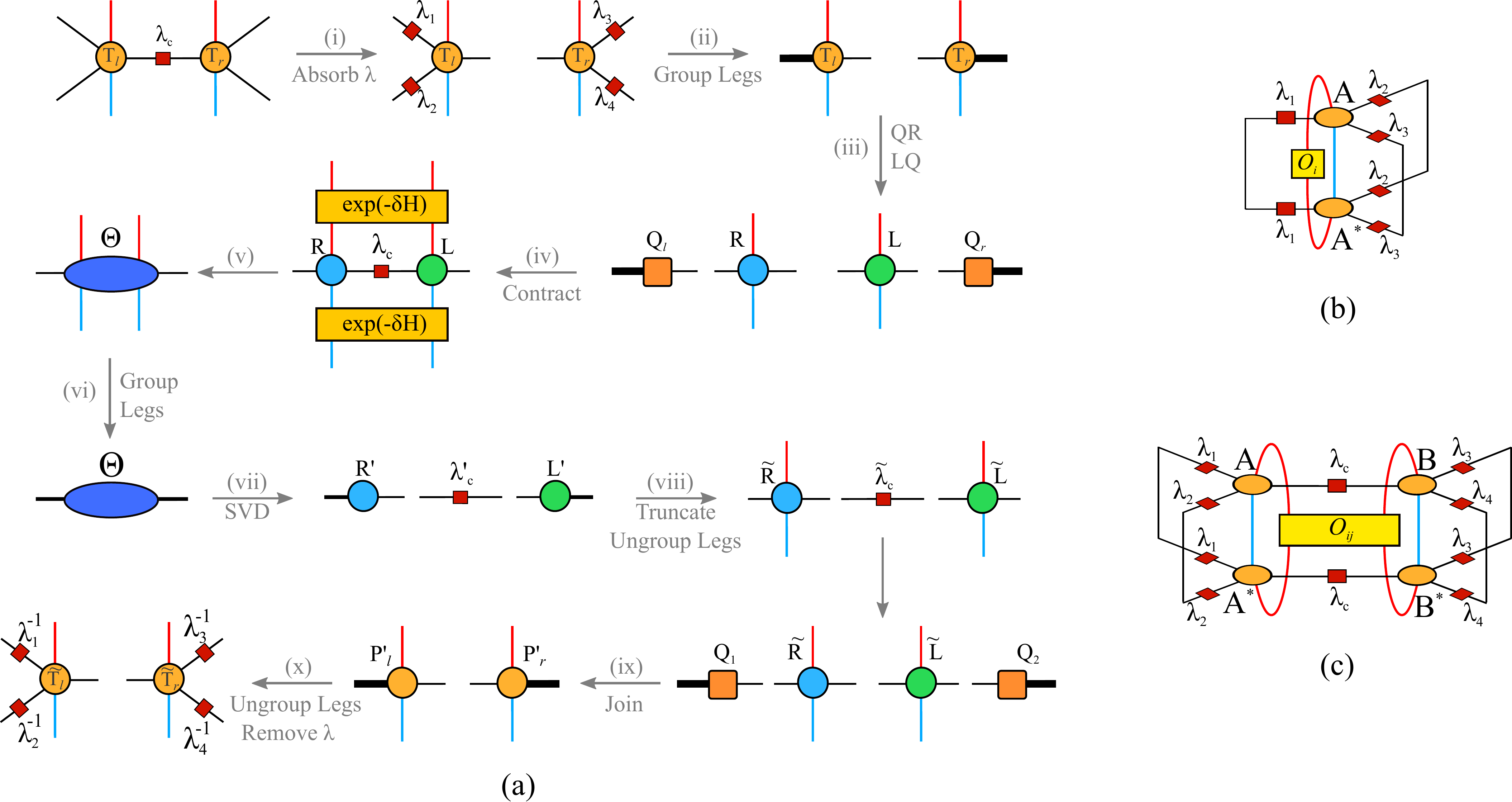}}
	\caption{(Color online) (a) Graphical representation of the SU optimization, used in the TgPEPS algorithm. In contrast to the gPEPS algorithm \cite{Jahromi2019} where the optimization is performed for ground state tensors with one physical leg, in the TgPEPS the optimization is performed over thermal PEPO tensors with two physical degrees of freedom. (b) One-site and (c) two-site expectation values, as computed with a mean-field environment, in the TgPEPS scheme. All these diagrams are for a 2d honeycomb lattice, but they can be straifgtforwardly generalized to any 3d structure, such as the ones considered in the main text.}
	\label{Fig:algorithm}
\end{figure*}

The thermal density matrix of the system is evaluated by iteratively applying $g$ on every shared link of the two neighbouring tensors $T_i,T_j$ and updating the tensors along the corresponding links. In this scheme, the update changes only the tensors along the link where a given gate is acting. Therefore, one can update lower-rank sub-tensors related to them and substantially reduce the computational cost of the algorithm \cite{Phien2015}, thus allowing to achieve larger bond dimension $D$. 

Let us briefly revisit how the SU proceeds for the sub-tensors, in the context of TgPEPS. Given a tensor network and its corresponding structure matrix, the SU consists of the following iterative main steps:
\begin{enumerate}
	\item Do for all edges $E_k$, $k\in[1,N_{Edge}]$ (columns of SM matrix) 
	\begin{enumerate}
		\item Find tensors $T_i,T_j$ and their corresponding dimensions connected along edge $E_k$. 
		\item Absorb bond matrices $\lambda_{m}$ to all virtual legs $m\neq k$ of $T_i,T_j$ tensors. 
		\item Group all virtual legs $m\neq k$ to form $P_l$, $P_r$ PEPO tensors. 
		\item QR/LQ decompose $P_l$, $P_r$ to obtain $Q_1$,$R$ and $L$, $Q_2$ sub-tensors, respectively \cite{Phien2015}.
		\item Contract the ITE gate $U_{i,j}$, with $R$, $L$ along the legs of both ket and bra together with $\lambda_k$ to form $\Theta$ tensor.
		\item Obtain $\tilde{R}$, $\tilde{L}$, $\tilde{\lambda}_k$ tensors by applying an SVD to $\Theta$ and truncating the tensors by keeping the $D$ largest singular values.
		\item Glue back the $\tilde{R}$, $\tilde{L}$, sub-tensors to $Q_1$, $Q_2$, respectively, to form updated PEPO tensors $P'_l$, $P'_r$.
		\item Reshape back the $P'_l$, $P'_r$ to the original rank-$(z+2)$ tensors $T'_i,T'_j$.
		\item Remove bond matrices $\lambda_{m}$ from virtual legs $m\neq k$ to obtain the updated PEPO tensors $\tilde{T}_i$ and $\tilde{T}_j$.
	\end{enumerate}
\end{enumerate}
Fig.~\ref{Fig:algorithm}-(a) shows all these steps graphically for the case of a 2d honeycomb lattice (the generalization to 3d is straightforward). By repeating the above ITE iterations $m/2$ times where $m.\delta=\beta$, the system will cool down to the desired temperature $T=1/\beta$. Once the TDM of the system at desired $T$ is obtained, we can calculate the expectation values of local operators as shown graphically for one- and two-body operators in Fig.~\ref{Fig:algorithm}-(b,c). 

In order to have an efficient and universal algorithm applicable to any infinite lattice, the following remarks are in order: (i) In steps (b), (c), (g) and (h) one can locate the lambda matrices corresponding to each leg of a tensor from rows of the SM. One can therefore design  clever functions for absorbing (removing) $\lambda$ matrices to (from) each tensor legs as well as for grouping (un-grouping) the non-updating tensor legs by using the information stored in each row of the SM. (ii) In our SU optimization, we perform the ITE iteration for $\delta=10^{-3}$. (ii) Furthermore, one can increase the stability of the SU algorithm by applying the gauge-fixing \cite{Jahromi2019}. 

Let us further note that the computational cost of the SU scales as $O(p^2 D^z)$, and evidently depends on the coordination number of the underlying lattice. Henceforth, the maximum achievable bond dimension $D$ is lattice dependent and is larger for structures with less coordination number, though structures with large $z$ usually need low $D$ because of entanglement monogamy.

\section*{Acknowledgements }

CPU time from ATLAS HPC cluster at DIPC is acknowledged.

\section*{Author contributions statement}

All authors contributed equally to the manuscript.

\end{document}